# The possibility of a Hall thruster operation in the absence of the anode sheath

L. Dorf [a), b)], V. Semenov [b)], Y. Raitses [a)] and N. J. Fisch [a)]

a) Princeton Plasma Physics Laboratory (PPPL), Princeton, NJ 08543
b) Institute of Applied Physics of Russian Academy of Science (IPFRAN), 46 Ulyanov St., Nizhnii Novgorod, 603000, Russia

A method of determining boundary conditions for quasi 1-D modeling of steady-state operation of a Hall Thruster with ceramic channel is presented. For a given mass flow rate and magnetic field profile the imposed condition of a smooth sonic transition uniquely determines plasma density at the anode. The discharge voltage determines the structure of the anode sheath and thus determines electron and ion velocities at the anode. These parameters appear to be sufficient for constructing a solution with given temperature profile. It is shown that a good correlation between simulated and experimental results can be achieved by selecting an appropriate electron mobility and temperature profile. The structure of the electrode sheath was studied theoretically over a wide range of input parameters, such as discharge voltage, incoming neutral velocity and channel length, and the possibility of realization of the no-sheath operating regime is discussed here.

## PROBLEM SETUP

Consider the conventional case of a HT with ceramic channel. The input parameters for any model of the HT are the experimentally controlled parameters, namely, the discharge voltage, $V_d$, the propellant mass flow rate, $\frac{dm}{dt}$, and the radial magnetic field profile, $B_r(z)$. We neglect the influence of the axial component of the magnetic field. The output parameters to be determined are the discharge current, $I_d$, the propellant utilization, and the profiles of ion velocity, ion density and potential.

To describe a steady-state operation of a Hall Thruster we consider the following physical processes. *Single ionization:* ions are born with the neutral velocity; *wall losses:* averaged over the channel cross-section; *ion acceleration:* toward the cathode, use hydrodynamic momentum equation for a mono-energetic ion flow with the ion velocity $V_i$; *closed electron drift:* azimuthal, in $E_z \times B_r$ direction; *electron diffusion:* toward the anode, with the electron flow velocity $V_e$; *free neutral motion:* assume the mono-energetic neutral flow with the constant neutral velocity $V_{a0}$. We also make a *quasineutrality* assumption: $n_i = n_e = n$, which is typical for a HT modeling.

Let us note that in most of our numerical simulations we used input parameters typical for the PPPL HT operation:[1] $V_d = 150 - 300V$, $\frac{dm}{dt} = 1.7 - 3.0 mg/s$ (propellant gas - Xenon), and $B_{max} \sim 130 Gs$. We used the analytical fit consisting of six gauss-functions for magnetic field profile near the channel median as $B_r(z)$ (Fig. 1):

In a 1-D description of a problem it is also necessary to select a distance from the anode, $L_c$, at which the voltage drop equals to $V_d$, in other words make a choice of the cathode plane. For now we just naturally choose it to be the plane, where the cathode tip is physically located, $L_c = 5.4 cm$ for PPPL HT, but we will later discuss this issue in a greater detail.

## GOVERNING EQUATIONS

In our quasi 1-D model all vectors are projected on to the $z$ - axis, where $z$ is the coordinate along the thruster axis, with $z = 0$ at the anode. The physical processes can be expressed mathematically as follows:

### A. Ion continuity equation

$$(nV_i)' = <\sigma V> n_a n - \frac{2}{H_{ch}} \times 0.55 n \sqrt{\frac{T_e}{M_i}} \times \Theta(L_{ch} - z), \quad (1)$$

where the prime sign denotes the derivative with respect to $z$. In the first term of equation (1), the ionization constant, $<\sigma V>(T_e)$, was obtained using experimental data for ionization cross-section, $\sigma_i^{Xe}(E_e)$.[2] The electron distribution function was assumed to be Maxwellian with the local temperature $T_e$, and then the analytical approximation for $<\sigma V>(T_e)$ was deduced. In the second term $H_{ch}$ and $L_{ch}$ are the width and the length of a channel respectively ($H_{ch} = 1.8 cm$, $L_{ch} = 4.6 cm$ for PPPL HT); theta function, $\Theta(L_{ch} - z)$, represents the absence of the wall losses outside of the channel; and the factor of two indicates the presence of two channel walls. The factor of 0.55 was obtained by solving the radial sheath problem in the hydrodynamic description with the ionization and without collisions, as described by Reimann in Ref [3], however not assuming the quasineutrality in the presheath.

### B. Ion momentum equation

$$(nV_iV_i)' = \frac{eEn}{M_i} - \frac{2}{H_{ch}} \times 0.55 n \sqrt{\frac{T_e}{M_i}} \times \Theta(L_{ch} - z) \times V_i + <\sigma V> n_a n V_{a0}, \quad (2)$$

where $E$ is the axial projection of the electric field and $n_a$ is the neutral density.

### C. Charge conservation

$$-nV_e + nV_i = J_d, \quad (3)$$

where $J_d = \frac{I_d}{eA_{ch}}$, $e$ is electron charge, and $A_{ch} = 40.7 cm^2$ is the channel cross-section.

### D. Electron momentum equation

$$en\mu_e^{-1} V_e = eEn + (n_e T_e)', \quad (4)$$

We describe the electron axial motion with the phenomenological electron momentum equation, (4), in which $\mu_e$ is the absolute value of the electron axial mobility in a radial magnetic field.[4] For the main purpose of this paper it is enough to assume Bohm diffusion, i.e. $\mu_e = \mu_e^{Bohm} = \frac{1}{16 B_r(z)}$. However, in order for numerical simulations to correlate well with the experiment, $\mu_e$ must be chosen more carefully, as we show later in this paper.

### E. Mass conservation

$$n_a V_{a0} + nV_i = J_{a0} + J_{i0}, \quad (5)$$

where $J_{a0}$ and $J_{i0}$ are neutral and ion fluxes at the anode respectively. We consider that no ions are coming out of the anode, and all ions hitting the anode recombine with electrons and return to the discharge as neutrals. We therefore obtain: $J_{a0} = J_m - J_{i0}$, where $J_m \overset{def}{=} \frac{dm/dt}{M_i A_{ch}}$ is the propellant flux. In simulations we consider a free molecular neutral flow out of the hot anode ($T_{anode} = 1000^0 C$) to obtain $V_{a0} = 113 m/s$.

### F. Electron energy equation

$$T_e(z) = Const \quad (6)$$

Let us first consider a case of constant electron temperature. It is known from experiments, that $T_e \sim 3-5 eV$ near the anode, and $T_e \sim 18-20 eV$ in the maximum of the temperature profile,[5] so we choose $T_e$ from that interval in our numerical simulations. Later in this paper we return to the question of determining the temperature profile.

BOUNDARY CONDITIONS

The above system of equations can be reduced to the system of two ordinary differential equations for density, $n(z)$, and ion flux, $J_i(z) = n(z)V_i(z)$:

$$\begin{cases} J_i' = n<\sigma V_e>(T_e)\dfrac{J_m - J_i}{V_{a0}} - 1.1\dfrac{n V_s}{H_{ch}} \\ n' = \dfrac{1}{1 - V_i^2/V_s^2}\left[\dfrac{e}{m_e\mu_e V_s}\dfrac{J_d - J_i}{V_s} - n\left((\ln T_e)' - \dfrac{<\sigma V_e>(T_e)}{V_s}\dfrac{J_m - J_i}{V_s}\right) - \right. \\ \qquad\qquad \left. - J_i\left(2\dfrac{<\sigma V_e>(T_e)}{V_s}\dfrac{J_m - J_i}{V_{a0}} - \dfrac{1.1}{H_{ch}V_s}\right)\right], \end{cases} \quad (7)$$

where $V_s = \sqrt{\dfrac{T_e}{M_i}}$ is the ion acoustic velocity. The $\Theta$-function in terms originating from the wall-losses term in (1) was omitted for simplicity.

If we specify the ion flux and the plasma density at the anode and the charge flux $J_d$, we can try to integrate (7) numerically. In other words, our system of equations contains 3 free parameters to be determined before the solution can be obtained: $n_0$, $M_0$ and $\overline{V}_0$, where $M = \dfrac{V_i}{V_s}$, $\overline{V} = \dfrac{V_e}{V_{te}}$, $V_{te} = \sqrt{\dfrac{T_e}{m_e}}$ is the electron thermal velocity, and the subscript "naught" means that functions are evaluated at the anode, $z = 0$. Let us point out, that $M_0$ and $\overline{V}_0$ explicitly enter in the charge flux, $J_d = n_0(-\overline{V}_0 V_{te0} + M_0 V_{s0})$, and the ion flux at the anode, $J_{i0} = n_0 M_0 V_{s0}$. Thus, we need 3 boundary conditions to provide the existence and uniqueness of the solution.

a. *Determining plasma density at the anode, $n_0$.*

Neglecting wall losses in the ion continuity equation and assuming $V_{a0} = 0$ in the ion momentum equation, in order to better demonstrate our approach to determining the free parameters, we can deduce the following normalized equation for ion Mach number:

$$d_t M = \dfrac{(1 + M^2)(1 - J_i/J_m)\beta - M^2(J_d/J_i - 1)\gamma}{1 - M^2}, \quad (8)$$

where $t \stackrel{def}{=} z/H_{ch}$, $\beta \stackrel{def}{=} \dfrac{<\sigma V>(T_e)J_m H_{ch}}{V_{a0}V_s}$, $\gamma \stackrel{def}{=} \dfrac{16\omega_{Be}H_{ch}}{V_s}$, and $\omega_{Be} = \dfrac{eB_r}{m_e}$ is electron cyclotron frequency.

The equation (8) describes the ion dynamics in quasineutral plasma. A similar equation describes the flow dynamics in the well-known de Laval nozzle.[6] The first (positive) term in the numerator of (8) originates from the ionization term in (1) and leads to ion acceleration in the subsonic region of the ion flow, i.e. where $M < 1$. The second (negative) term in the numerator of (8) is originally the electric field term from the equation (2), and it effectively works in subsonic plasma as an ion drag. The denominator appears essentially because of the electron pressure and it turns to zero at the boundary of the subsonic flow, at which $M=1$. This leads to a singularity, typical for quasineutral plasmas and called the "sonic transition".[6] We look only for a non-singular solution of our system which describes a smooth behavior of all physical values in the vicinity of the sonic transition point, $z_{st}$, at which $M = 1$. Let us point out that Fruchtman and Fisch in Ref. [7] considered the possibility of abrupt sonic transition in HT with an additional electrode placed inside the channel, and Ahedo et al in Ref. [8] proposed a "choked-exit" type of solution, in which ions reach the sound velocity right at the channel exit. However, all of the authors considered a smooth sonic transition in their models of the conventional Hall Thrusters.

As can be seen from (8), in order for the sonic transition point to be regular it is necessary that the drag and acceleration terms are equal at this point. Both of these terms depend on $J_d$ and $J_{i0}$, i.e. on the free parameters that we set at $z=0$. Out of the 3 free parameters, only $n_0$ and $M_0$ enter explicitly in both $J_d$ and $J_{i0}$, and, as will be shown in the next paragraph, $M_0$ and $\overline{V}_0$ are physically interdependent. So, we conclude that exactly the choice of $n_0$ is responsible for the smooth sonic transition. We set $M_0$ and $\overline{V}_0$ in the interval from 0 to 1 and tried to select $n_0$ numerically in order to obtain a non-singular (NS) solution. It was shown by a comprehensive scanning over all reasonable for HT values of $n_0$, that a smooth sonic transition takes place only if $n_0$ equals to a certain *unique* value, $n_0^{NS}$, which depends, of course, on $(\overline{V}_0, M_0)$. If $n_0 > n_0^{NS}$, the drag term appears to be too big and $M$ does not reach 1 anywhere in the channel; and if $n_0 > n_0^{NS}$, the drag term is too small and a numerator in (8) appears to be greater than zero at $z_{st}$, which leads to a singularity (Fig. 2). Thus, for given $(\overline{V}_0, M_0)$ the requirement of the sonic transition point to be regular results in the unique value of $n_0$.

b.  *Determining electron and ion velocities at the anode, $\overline{V}_0$ and $M_0$.*

There are two possibilities in determining $\overline{V}_0$ and $M_0$. For the same discharge voltage the Hall thruster as every gas discharge may operate in one of the two regimes - with and without the anode sheath. If there is a sheath then, like in Ref [8], we obtain that $M_0 = -1$. The electron velocity, $\overline{V}_0$, in this case must be selected in order to obtain a total voltage drop in plasma equal to a given $V_d$: $\int_{0+}^{Lc} E(z)dz = V_d$, where $L_c$ is the distance from the anode to the cathode plane, and "$0+$" means that the integration must be produced only over quasineutral plasma (we neglected a sheath voltage drop here, because $T_e$ at the anode is usually very small in a real HT). If there is no sheath and plasma is quasineutral up to the anode, then $\overline{V}_0 = -\overline{V}_{max}$, where $\overline{V}_{max}$ is determined only by the electron distribution function at the anode (we used $\overline{V}_{max} = 0.4$ in our simulations, assuming Maxwellian distribution). In this case ions accelerated in the presheath toward the anode do not reach the sound velocity and $M_0$ is determined by a given $V_d$.

To resolve this indeterminacy we numerically scanned in the $(\overline{V}_0, M_0)$ plane along the physically possible curve, as shown on Fig. 3.a. We found that $V_d$ and $I_d$ monotonically grow as we gradually transfer from "sheath" to "no sheath" regime (Fig. 3.b). For $V_d > V_d^*$, where $V_d^*$ correspond to the point $(-\overline{V}_{max}, -1)$ in the $(\overline{V}_0, M_0)$ plane, there is no anode sheath. So, for given $T_e$ and $\mu_e$ the discharge voltage, $V_d$, uniquely determines the operating regime and $(\overline{V}_0, M_0)$. The boundary condition issue is resolved.

The fact that for a given electron temperature the discharge voltage uniquely determines the plasma flow near the anode makes the Hall thruster analogous to a well-known Langmuir probe (Fig. 3.c). However, the similarity between the role of anode in HT and the role of Langmuir probe in plasma is not exact and can be used only for illustrative purposes. The change of $V_d$ in a HT leads to a change of the plasma potential in the near anode region (where $B_r(z)$ is weak) and to a change of $(\overline{V}_0, M_0)$; the latter in turn leads to a change of the plasma density at the "anode" (again, in a HT we are only considering quasineutral plasma). At the same time the change of the probe bias only changes the sheath–presheath structure and does not change physical quantities in the quasineutral plasma. Besides, the probe is usually placed in the plasma body far from the walls, so, unlike the HT, there are no wall losses in the probe problem. Thus, the specific calculation of $V_d$ and $I_d$ as we move from "Sheath" to "No sheath" region in the $(\overline{V}_0, M_0)$ plane was necessary and the indeterminacy described above could not have been resolved just by using an analogy with the probe.

## SOLUTION

We used the above boundary conditions to determine free parameters and obtain a solution with several constant temperatures. It appeared that, like in some other models,[7] at large temperatures all of the propellant is ionized in a very short region near the anode, and if we choose smaller temperatures, we get the propellant utilization, $\frac{J_i(L_{ch})}{J_m}$, atypically small for a HT. It is interesting to notice that at temperatures smaller than a certain lower threshold, ionization appears to be insufficient for normal operation of the thruster and it becomes impossible to build a non-singular solution with supersonic ion velocity at the thruster exit with any free parameters.

However, we showed that the same approach to determining free parameters can be applied and the solution can be constructed in the case of any given shape of temperature profile, qualitatively similar to experimental,[5] if maximal temperature, $T_{max}$, is chosen to be large enough. Then we comprehensively investigated the dependence of the solution on a shape of the temperature profile, and for each considered point [$V_d$, $J_m$, $B_r(z)$] in the typical PPPL HT operational range were able to determine $T_e(z)$, that allows us to obtain the experimental value of $I_d$, and the ratio $\frac{n_{max}}{n_0} \sim 10$, which is typical for HT.[5] From the same argument we have determined the actual value of electron mobility: $\mu_e \sim (1/8 - 1/6) \mu_e^{Bohm}$, which tends to increase with the increase of the discharge voltage. The fact that electron mobility in Hall thrusters appears to be several times less than the one obtained with a Bohm diffusion concept was also discovered by Keidar et al,[9] and some other authors.[10,11]

The numerically obtained profiles are shown on Fig. 4. The propellant utilization (about 86%) and potential profile were found to be in a very good agreement with experiment.[12]

As was already mentioned, the determination of a cathode plane location requires an additional discussion. From experimental measurements with the PPPL HT we know that the electric field goes to zero and the electric potential saturates at several centimeters from the thruster (beyond the cathode), so that a voltage drop equal to the discharge voltage occurs only between the anode and any plane located in the saturation region. In all our simulated solutions we also obtained that saturation region on the potential profile for $z > L_{sat}$, where $L_{sat}$ depends on [$V_d$, $J_m$, $B_r(z)$], and for all operational points it appeared that $L_{sat} > L_c$. So it seems very natural for quasi 1-D modeling to use the condition that a voltage drop equal to $V_d$ occurs between the anode, $z = 0$, and any point in the saturation region, $z > L_{sat}$, instead of $\int_{0+}^{Lc} E dz = V_d$, in other words, choose the cathode plane at infinity. We showed that the same approach to determining free parameters can be applied for constructing a solution with cathode plane located at any distance from the anode, up to infinity. However, outside of the channel a quasi 1-D description, which does not take into account radial divergence of the plasma jet, the change in electron mobility and other possible effects of the absence of channel walls, is the less adequate, the further from the channel exit we are trying to use it. As a result, the best correspondence between simulated and experimental results was still achieved with $L_c$ chosen in the plane, in which the tip of the actual cathode is physically located. Although in this problem setup, as could be expected, the saturation region of the potential profile (beyond the cathode plane) looks different from the one observed in experiments.

## DISCUSSIONS

For most of the considered ($V_d, J_m$) in the typical PPPL HT operational range the anode sheath appeared to take place, so we suggest that $M_0 = -1$ can be used as a universal BC for normal operation (moderate discharge voltages) of a HT. As was shown, $\overline{V}_0$ in this case must be selected in order to obtain a desired $V_d$. The negative ion flux toward the anode was indeed measured in experiments.[5] But for discharge voltages greater than a certain value sheath indeed disappeared and in order to obtain a solution we had to use an alternative, "no sheath" type boundary conditions: $\overline{V}_0 = -\overline{V}_{max}$, $M_0$ is determined by $V_d$. We also found that for the same $V_d$, $J_m$ and $B_r(z)$ an absolute value of $\overline{V}_0$ increases and may even reach $\overline{V}_{max}$ when we artificially decrease the channel length from the anode side. The "no sheath" regime was also observed experimentally.[13] Thus, we want to emphasize that "no sheath" type of BC is not just a physical abstraction; these BC indeed appear to be relevant for modeling of certain configurations and regimes of operation of a HT.

As was mentioned, in order for the described approach to determining free parameters to work properly for discharge voltages in the practically used range, $T_{max}$ for the electron temperature profile, $T_e(z) = T_{max} \cdot Shape(z)$, must be chosen large enough for a given $\mu_e$. Otherwise, as we move in the $(\overline{V}_0, M_0)$ plane along the physically possible curve from "Sheath" to "No sheath" region (Fig. 3.a), we will find the discharge current to very slowly increase, whereas the discharge voltage will significantly decrease. Of course, in the real HT the decrease of discharge voltage at the same magnetic field profile leads to the decrease of a discharge current and, as was shown before, the decrease of $V_d$ should occur when we move from "No sheath" to "Sheath" region, not on the contrary. We showed that the minimal value of $T_{max}$ at the same mass flow rate depends strongly on the incoming neutral velocity. Basically, if we know $T_{max}$, for which solution can be constructed with a certain $V_{a0}$, then in order to be able to construct a solution with another $V_{a0}$, we must choose $T_{max}$ so that to keep the ratio $\frac{<\sigma V>(T_{max})}{V_{a0}}$ approximately the same.

It is interesting to notice that our approach to determining the free parameters involves only global physical phenomena: anode sheath and sonic transition. So, we find this approach applicable for 2-D modeling, especially considering that 2-D effects take place mainly in the region of a strong magnetic field, far from the anode. Let us also emphasize that we showed an applicability of the described approach to determining free parameters for modeling of Hall thrusters with different profiles of the magnetic field, qualitatively similar to the one used in a PPPL HT. And we also believe that the same approach can be applied for modeling of other types of Hall thrusters such as *segmented electrode HT* and *anode layer thruster*.[1, 14]

ACKNOWLEGEMENT

The authors would kindly like to thank D. Staack, A. Smirnov and A. Litvak for very useful discussions and their comments on this paper.

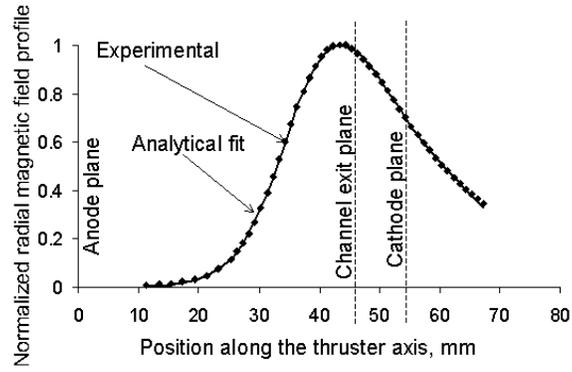

Fig. 1. A typical magnetic field profile for the PPPL HT

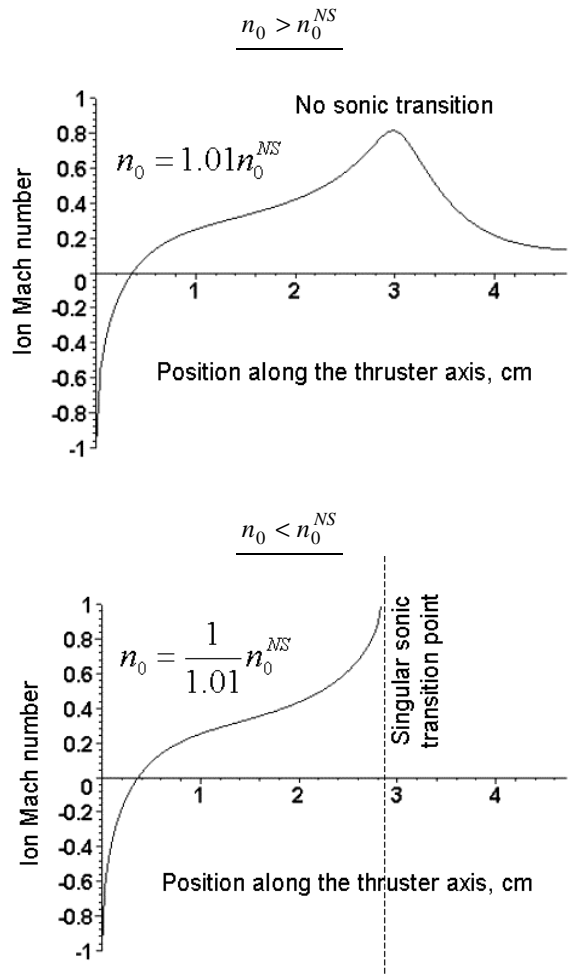

Fig. 2. Types of the ion velocity spatial behavior for different values of the plasma density at the anode. Case $T_e = Const$

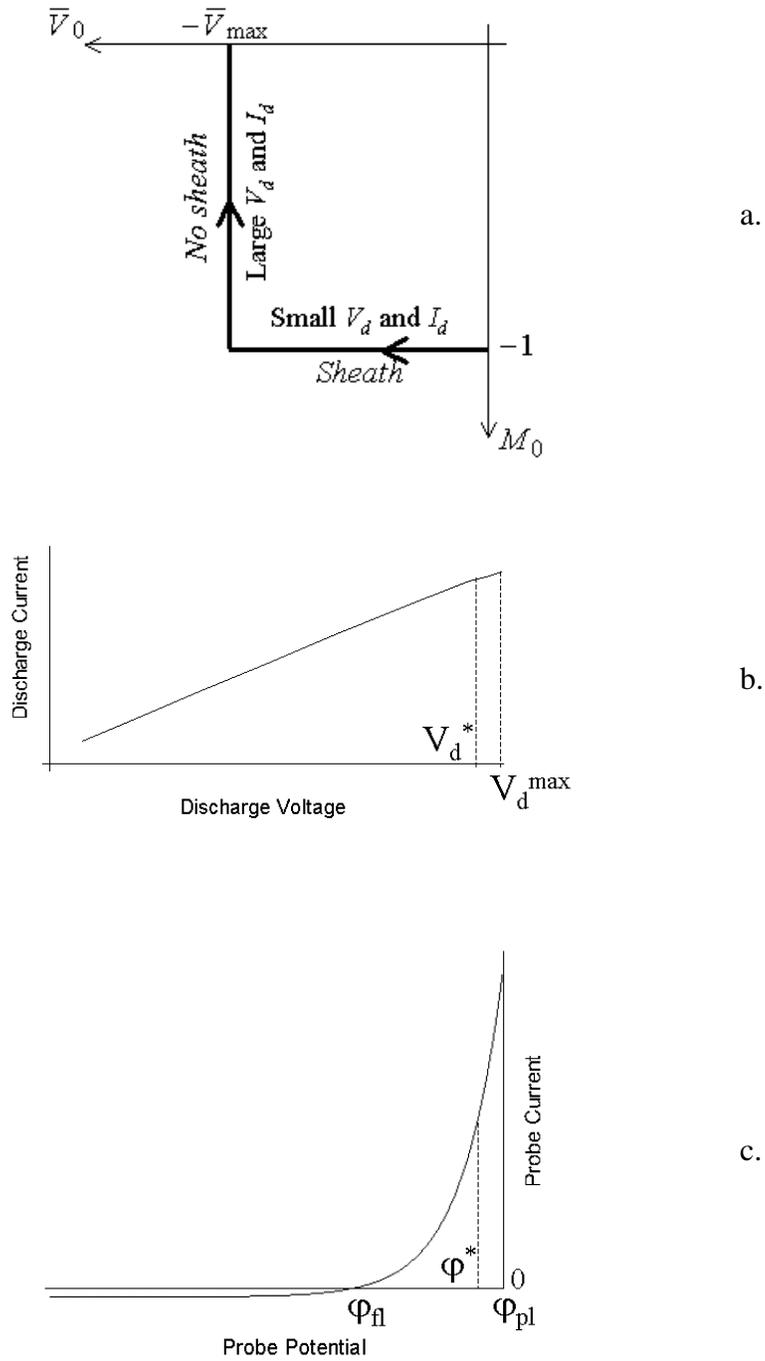

Fig. 3. a. A physically possible curve in the $(M_0, \overline{V}_0)$ cross-section of the free parameters space.
b. A schematic picture of the Hall thruster V-I characteristic. $V_d^*$ and $V_d^{max}$ correspond to the points $(-\overline{V}_{max}, -1)$ and $(-\overline{V}_{max}, 0)$ in the $(\overline{V}_0, M_0)$ plane, respectively; for $V_d > V_d^*$ there is no anode sheath.
c. A schematic picture of the Langmuir probe V-I characteristic. $\varphi^*$ separates "sheath" and "no sheath" cases. $\varphi_{fl}$ and $\varphi_{pl}$ are floating and plasma potentials respectively.

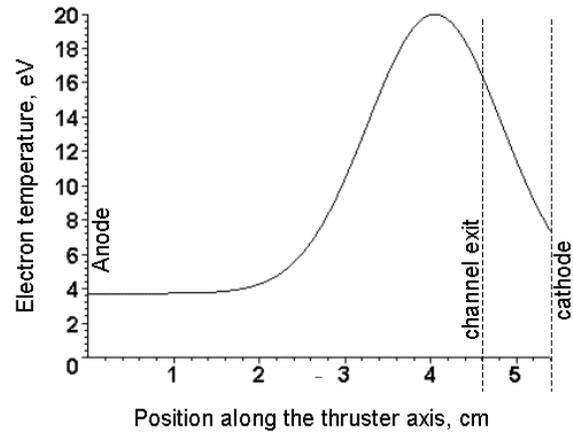

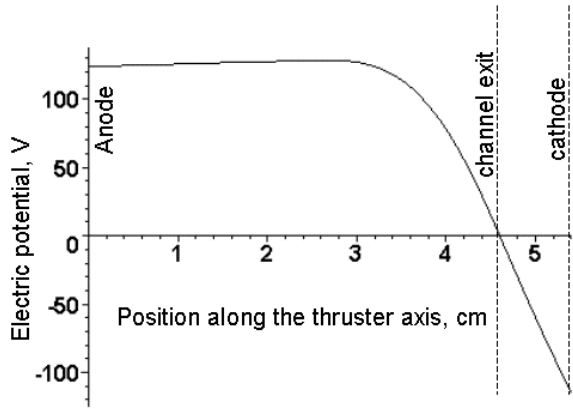

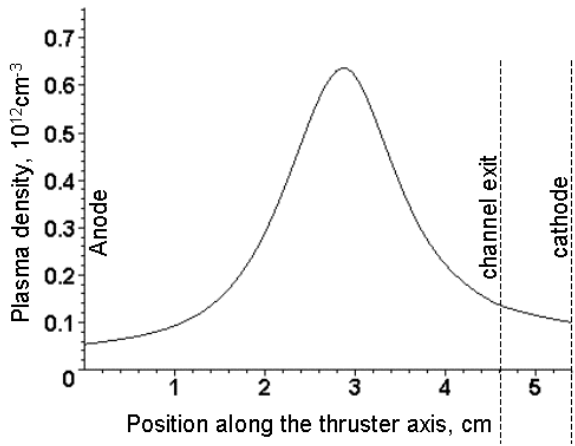

Fig. 4. The numerically obtained profiles in a Hall thruster. For $V_d = 240V$, $\frac{dm}{dt} = 1.7 mg/s$. Zero potential was chosen at the channel exit. $z_{st} = 3.52 cm$, $I_d = 1.63A$